\documentclass[11pt]{aastex}
\usepackage{graphicx,emulateapj5,apjfonts}
\usepackage{onecolfloat}
\usepackage{epsf}

\shortauthors{Bell et al.}
\shorttitle{On the decline of the cosmic SFR}

\newcommand{\ha}{{\rm H$\alpha$ }}

\newcommand{\combo}{{\rm COMBO-17 }}

\slugcomment{{\sc The Astrophysical Journal; to appear 1 June 2005 }}

\begin{document}

\def\head{
\title{Towards an understanding of the rapid decline of the cosmic star
formation rate}

\author{Eric F.\ Bell$^1$, 
Casey Papovich$^2$, Christian Wolf$^3$, 
Emeric Le Floc'h$^2$, John A.\ R.\ Caldwell$^{4,5}$, 
Marco Barden$^1$, Eiichi Egami$^2$, Daniel H.\ McIntosh$^6$, 
Klaus Meisenheimer$^1$, Pablo G.\ P\'erez-Gonz\'alez$^2$, G.\ H.\ Rieke$^2$, M.\ J.\ Rieke$^2$, Jane R.\ Rigby$^2$, and Hans-Walter Rix$^1$}
\affil{$^1$ Max-Planck-Institut f\"ur Astronomie,
K\"onigstuhl 17, D-69117 Heidelberg, Germany; 
\texttt{bell@mpia.de} \\ 
$^2$ Steward Observatory, The University of Arizona, 933 North 
Cherry Avenue, Tucson, AZ 85721, USA \\
$^3$ Department of Physics, Denys Wilkinson Building, University of 
Oxford, Keble Road, Oxford OX1 3RH, UK \\
$^4$ Space Telescope Science Institute, 3700 San Martin Drive, 
Baltimore, MD 21218, USA \\
$^5$ Present address: University of Texas, McDonald Observatory,
     Fort Davis, TX 79734, USA \\
$^6$ Department of Astronomy, University of Massachusetts,
710 North Pleasant Street, Amherst, MA 
01003, USA
}

\begin{abstract}

We present a first analysis of deep 
24$\mu$m observations with the {\it Spitzer Space Telescope} 
of a sample of 
nearly 1500 galaxies in a thin redshift 
slice, $0.65 \le z < 0.75$. We combine
the infrared data with redshifts, rest-frame 
luminosities, and colors from COMBO-17, and
with morphologies from {\it Hubble Space Telescope} images collected by the 
GEMS and GOODS projects. To characterize the 
decline in star-formation rate (SFR) since 
$z \sim 0.7$, we estimate the total thermal infrared (IR)
luminosities, SFRs, 
and stellar masses for the galaxies in this sample.
At $z \sim 0.7$, nearly 40\% of intermediate and high-mass 
galaxies (with stellar masses 
$\ge 2 \times 10^{10}$ M$_\odot$) are undergoing a period of intense star 
formation above
their past-averaged SFR. 
In contrast, less than 1\% of equally-massive galaxies in the local 
universe have similarly intense star formation activity.
Morphologically-undisturbed galaxies dominate 
the total infrared luminosity density and SFR
density: at z $\sim$ 0.7, more than half of the
intensely star-forming galaxies have spiral
morphologies, whereas less than $\sim$ 30\% are strongly 
interacting.  Thus, a decline in major-merger rate is 
not the underlying cause of the rapid decline in 
cosmic SFR since $z \sim 0.7$.  Physical 
properties that do not strongly affect galaxy morphology --- 
for example, gas consumption 
and weak interactions with small 
satellite galaxies --- appear to be responsible.

\end{abstract}

\keywords{galaxies: evolution --- galaxies: general --- 
galaxies: stellar content --- galaxies: fundamental parameters --- galaxies: interactions --- infrared: galaxies}
}

\twocolumn[\head]

\section{Introduction}

\vspace{0.3cm}

It has become clear in the last decade
that the average star-formation rate (SFR) per unit comoving
volume --- the so-called cosmic SFR --- has
declined by an order of magnitude since $z \sim 1$
\citep{lilly96,madau98,hogg98,flores99,haarsma00,hopkins04}.  
A number of physical processes
may contribute to this decline: e.g., a declining rate
of major galaxy mergers, a drop in the rate of 
minor tidal interactions, or the progressive consumption of cold gas.  
Yet, many empirical aspects of this declining cosmic
SFR are frustratingly unclear.  Does the 
star formation in disks evolve slowly, while merger-induced
starbursts evolve rapidly?  Does the decreasing number density of 
galaxies with high SFR reflect a decline in SF 
in Milky-Way mass galaxies or reflect a disappearance
of a population of bursting dwarf galaxies?  
In this paper, we present a first basic analysis of deep 24{\micron}
observations taken by the MIPS Team
\citep{rieke04,papovich04,rigby04}.  We combine these infrared (IR) data
with the COMBO-17
redshift and spectral energy distribution (SED) 
survey \citep{wolf04} and the largest
existing {\it Hubble Space Telescope} ({\it HST}) ACS mosaic 
\citep[GEMS; the Galaxy Evolution 
from Morphology and SEDs; ][]{rix04} to study nearly 1500 galaxies 
with $0.65 \le z < 0.75$ in the 30$' \times 30'$ 
extended Chandra Deep Field South (CDFS), with the goal of exploring 
the morphologies, stellar masses and SFRs of 
galaxies, and probing the physics driving the decrease
in the cosmic SFR to the present day.

\subsection{Understanding the form of the cosmic 
star formation history}

The basic form of the cosmic star-formation history (SFH)
was initially established using observations of rest-frame
ultraviolet (UV) luminosity \citep{lilly96,madau96,madau98,steidel99}, 
which in the absence of dust
is an excellent indicator of the bolometric light 
output from young, massive stars.  As rest-frame UV is redshifted
into the optical, it becomes observationally straightforward to 
establish a deep census of unobscured young stars, well below
the knee of the luminosity function.  Yet, studies of both local 
and distant galaxies have shown that only a small fraction 
of the UV photons from $\sim L^*$ galaxies
escape unhindered \citep[dwarf galaxies are often optically
thin to UV photons, but do not significantly contribute
to the SFR density;][]{wang96,adel00,bellsfr}.  Thus,
the observed drop in UV luminosity density may not track 
quantitatively the drop in cosmic SFR.

Mindful of the limitations of rest-frame UV, other SFR indicators
have been explored.  Rest-frame optical line 
emission reflects the ionizing photon output from very massive young stars, 
in the absence of dust extinction
\citep[see][for a detailed discussion]{k98}.  Emission line-derived
SFRs have been studied intensively in the local universe
\citep{gallego95,brinchmann04}, and have been 
probed out to $z \sim 3$ for the very brightest galaxies 
using rest-frame O{\sc ii} and Balmer line 
emission \citep{hogg98,yan99,glaze99,tresse02,hipp03,erb03}. 
These studies find a considerable drop in the average
SFR per unit volume in luminous galaxies since $z \sim 1$, 
but owing to the difficulty of measuring Balmer decrements
of high redshift galaxies, considerable uncertainties remain.

Radio and X-ray emission from distant galaxies also
constrains their SFRs \citep{condon92,cohen03}.
Observational constraints limit the 
applicability of both indicators to just the brightest
sources at this stage; a substantial drop in the volume-averaged
SFR in very luminous sources is again seen since $z \sim 1$
\citep[e.g.,][]{haarsma00,cohen03}.  Yet, both X-ray and radio luminosities
are strongly boosted by the presence of an active galactic
nucleus (AGN), and the physical origin of both is not well-understood.

One of the most intuitive SFR indicators is the luminosity 
of a galaxy in the thermal IR.  For most star-forming
galaxies, the thermal IR emission is dominated by reprocessed
UV photons from massive young stars that are dust-enshrouded
\citep[see][for an excellent review]{k98}.
For most galaxies the combined IR and UV luminosity is a relatively
robust indicator of the bolometric output from young stars, and 
therefore a good proxy for SFR 
\citep[e.g.,][]{fluxrat,buat02,bellsfr}, especially if the modest
contribution to the IR flux from dust-reprocessed light from older
stellar populations is accounted for 
\citep[typically $\sim 30$\% or less;][]{misiriotis01,bellsfr}. 
Until recently, instrumental sensitivity has been a key limitation
at these wavelengths;  only galaxies with
IR luminosities $\ga 10^{11} L_{\sun}$ 
have been observed at cosmologically-interesting redshifts by 
the Infrared Space Observatory (ISO).  Through detailed
modeling and analysis of these observations, it is clear that 
highly-luminous IR galaxies were much more common at 
$z \sim 1$ than they are today
\citep[e.g.,][]{flores99,elbaz02,mann02,pozzi04}.
Observations of sub-mm emission from distant, 
intensely star-forming galaxies have extended 
this picture to $z \ga 2$ \citep[e.g.,][]{blain99}. 
Yet, despite this encouraging progress, it is still unclear
how the evolution of these highly-luminous galaxies 
is linked to the much more numerous fainter galaxies that dominate
the volume-averaged SFR.  

\subsection{Understanding the physical drivers
of star formation}

A wide range of processes may contribute to the order-of-magnitude drop
in cosmic SFR since $z \sim 1$: major mergers, tidal
interactions, gas consumption, ram-pressure stripping, 
galaxy harassment, or the secular evolution of bars.
Deciding which processes dominate involves a number 
of observational and interpretive challenges.  As an example,
it is challenging to measure the major merger rate robustly; 
close pair statistics suffer from 
contamination, automated asymmetry measures are
sensitive to contamination
from irregular galaxies and very close projected pairs, and 
visual morphological classifications are subjective, and 
less sensitive to prograde interactions owing to their lack of 
pronounced tidal tails.  

There is evidence that these different physical processes
play a role in shaping the SFRs of galaxies.  Many merging
galaxies form stars intensely \citep[e.g.,][]{sanders96}.  Given 
the observational evidence for a drop in the merger rate since $z \sim 1$
\citep[e.g.,][]{lefevre00,patton02,conselice03},
it is indeed plausible that the decline in the volume-averaged cosmic SFR
is driven at least in part by this drop in merger rate.
Furthermore, minor interactions seem to enhance SFRs, at least
in the local universe \citep[e.g.,][]{barton00}, plausibly by 
exciting unstable modes such as bars or strong transient spiral
arms in the affected galaxies.
The exhaustion of cold gas could also play a role.
Indeed, most massive nearby galaxies have low gas fractions $\la 10$\%
\citep[e.g.,][]{ke94,bdj00}, and the current SFRs of many local galaxies will
exhaust the available gas supply in $\sim 5$ Gyr \citep[e.g.,][]{ke94}.
It is possible that fresh infall or recycled gas can help to 
sustain star formation for a longer time.  Yet, it is also possible
that we live in a special epoch, when the star formation in 
massive galaxies is switching off.  

\subsection{The object of this paper}

This paper is one of a series in which we explore different
aspects of the declining cosmic SFR.  
\cite{wolfuv} study the contribution of different
galaxy morphologies to the rest-frame 2800{\AA} luminosity density
for rest-frame $B$-band selected galaxies 
in a thin $0.65 \le z < 0.75$ redshift slice from COMBO-17 and GEMS.  In this 
paper, we supplement this 
optically-selected sample with 24{\micron} data from 
Spitzer, to explore the 
contribution of obscured star formation\footnote{An additional
advantage of adopting such a narrow redshift slice for this paper is
that the effects of {\it differential}
$(1+z)^4$ surface brightness dimming, morphological
$k$-corrections, and uncertainties in 
the conversion of 24{\micron} flux to total IR
are minimized.}.  Le Floc'h
et al. (in preparation) explore the IR luminosity function over 
the range $0 < z < 1$
using MIPS 24{\micron} data and COMBO-17 redshifts. 
Papovich et al. (in preparation) explore the morphologies of 
the galaxies that dominate the turn-over in the 24{\micron}
number counts, as a clue to understanding the drivers of IR-obscured star
formation.  

In this paper, we study the MIPS 24{\micron} properties of a complete 
sample of rest-frame $B$-band selected 
$0.65 \le z < 0.75$ galaxies in the CDFS\footnote{At this stage, 
we cannot and do not study completely obscured galaxies with 
extremely faint or non-existent UV/optical detections.  
Preliminary indications are that these optically faint
IR sources are optically faint largely owing to their relatively high redshift
\citep[$z \ga 1$; e.g.,][]{papovich04} rather than having a lower
redshift and higher obscuration.  See Appendix \ref{app:c17comp}, 
where this issue is explored in more detail.}.
Owing to the high sensitivity and relatively
tight point spread function of 24{\micron} {\it Spitzer} observations,
many more star-forming galaxies are detected at 24{\micron} than at 
70{\micron} and 160{\micron}.
We use IR template spectra, tuned to reproduce the whole range of local 
galaxy IR SEDs, to transform 24{\micron} luminosity into 
total IR luminosity.  
For the smaller area imaged by Chandra to date, 
we incorporate the Chandra X-ray data \citep{alex03} 
to explore the importance of AGN at
24{\micron} \citep[see also][]{rigby04}.  Our main goals are to identify
all $z \sim 0.7$ galaxies with high SFRs, explore their morphologies, 
and characterize the importance
of massive starbursting galaxies compared to the present-day 
universe, as a probe of the physics driving the declining cosmic SFR.

The paper is set out as follows.
In \S \ref{sec:data} we discuss the data.  In \S \ref{sec:ana} we 
outline the methods used to estimate total IR flux, SFR and 
stellar mass, and the main sources of uncertainty.  In 
\S \ref{sec:iruv} we present 
trends in the IR/UV ratio.  In \S \ref{sec:res1} we explore the 
contribution of different morphological types to the total SFR
at $z \sim 0.7$.  In \S \ref{sec:massive} we discuss
intense star formation in massive galaxies, and how this changes over 
the last 7 Gyr.  In \S \ref{sec:disc} we compare with existing data,  
explore the contribution of AGN, compare with theory, 
and discuss future improvements 
in this analysis.  In \S \ref{sec:conc} we present our conclusions.
Appendices \ref{app:c17comp} and \ref{app:goods} discuss possible
incompleteness in our optically-selected catalog, and uncertainties
in the galaxy morphological classifications.
Throughout, we express the estimated integrated UV and IR luminosities 
in terms of the  
bolometric luminosity of the sun $L_{\sun} = 3.9\times10^{26}$W.  
We assume $\Omega_{\rm m} = 0.3$, 
$\Omega_{\Lambda} = 0.7$, and $H_0 = 70$\,km\,s$^{-1}$\,Mpc$^{-1}$
\citep{spergel03}.  Estimates of SFR and stellar mass assume a
universally-applicable \citet{kroupa01} IMF\footnote{
In the case where the stellar IMF varies as a function of time and/or
from galaxy-to-galaxy, neither SFRs nor stellar masses will be 
robust except perhaps
at the order-of-magnitude level.}.

\section{The Data} \label{sec:data}

\subsection{MIPS Data}

Spitzer observed a $1{\arcdeg} \times 0{\fdg}5$ field around the CDFS
in January and February 2004 as part of the time allocated to the
Spitzer Guaranteed Time Observers (GTOs).  The MIPS 24{\micron} data
was taken in slow scan--map mode, with individual exposures of 10~s.
We reduced the individual image frames using a custom data--analysis
tool (DAT) developed by the GTOs \citep{dat}\footnote{To date, our preliminary tests indicate that the
DAT reduction of MIPS 24{\micron} data is nearly identical to that
provided by the Spitzer Science Center pipeline.}.  The reduced images
were corrected for geometric distortion and combined to form full
mosaics.  The final mosaic has a pixel scale of
$1\farcs25$~pixel$^{-1}$ and an image PSF FWHM of $\simeq 6$\arcsec.
The Spitzer astrometry is aligned to 
the ESO Imaging Survey \citep[EIS;][]{arnouts01} with 
a typical accuracy $\sim$0.5\arcsec.  The overall astrometric solution 
is then shifted by nearly 1{\arcsec} 
by matching to the much sparser Two Micron All Sky Survey
catalog (2MASS).  This revised astrometric solution is 
matched to $\la 0.1$\arcsec
with the COMBO-17 coordinate system.

Source detection and photometry were performed using techniques
described in \citet{papovich04}; based on the analysis in that
work, we estimate that our source detection is 80\% complete at
83~$\mu$Jy in the 24{\micron} image.  Based on our noise
estimates, the photon noise in the 24{\micron} image
is roughly equal to the confusion noise at this depth
\citep[see, e.g.,][]{dole04}.

\subsection{Optical Data: \combo and GEMS}

To date, \combo has surveyed three disjoint
$\sim 34' \times 33'$ southern and equatorial fields
to deep limits in 5 broad and 12 medium passbands.
Using these deep data in conjunction with 
non-evolving galaxy, star, and AGN template spectra, objects
are classified and redshifts assigned for $\sim 99$\% of the
objects to a limit of $m_R \sim 23.5$.  Typical galaxy redshift accuracy
is $\delta z/(1+z) \sim 0.02$ \citep{wolf04},
allowing construction of $\sim 0.1$ mag accurate 
rest-frame colors and absolute magnitudes (accounting for distance
and $k$-correction uncertainties).  Astrometric accuracy is 
$\sim 0.1$\arcsec.

To explore galaxy morphology in the rest-frame
optical from a single observed passband, we
study galaxies in one thin redshift slice.   
Here, we select galaxies from \combo in the 
CDFS for morphological classification in the narrow interval
$0.65 \le z < 0.75$ (corresponding to $\Delta t \sim 0.5$\,Gyr, minimizing
galaxy evolution across this slice).  
At this redshift, F850LP samples roughly rest-frame
$V$-band, and Spitzer's 24{\micron} passband 
samples rest-frame 14{\micron}, allowing comparison with 
local samples.  Furthermore, $z \sim 0.7$ is close to the median
redshift of COMBO-17, maximizing the sample size.

We use F850LP imaging from the GEMS
survey \citep{rix04} to provide 
$0.07"$ resolution rest-frame $V$-band data for 1492 
galaxies with $0.65 \le z < 0.75$.
GEMS surveys a $\sim 28' \times 28'$
portion of the CDFS in the F606W and F850LP passbands
to deep limits using the Advanced Camera for Surveys \citep{ford03}
on the {\it HST}.  The GEMS area is covered by a multiple, overlapping
image mosaic that includes the smaller but deeper GOODS
area \citep{giavalisco04}.  One orbit per pointing
was spent on each passband (63 GEMS tiles and 15 GOODS
tiles), allowing galaxy detection to a limiting surface 
brightness within the half-light radius 
of $\mu_{\rm F850LP,AB} \sim 24$\,mag\,arcsec$^{-2}$
\citep{rix04}.  At $z = 0.7$, ACS resolution 
corresponds to $\sim 500$\,pc resolution, 
roughly equivalent to $\sim 1''$ resolution at Coma
cluster distances.

Galaxy classification was carried out by-eye
on the $z \sim 0.7$ sample by 
EFB, CW and DHM using the GEMS
F850LP imaging\footnote{Automated
classification using profile fitting is highly consistent
with the by-eye classifications that we adopt here, in terms
of differentiating between bulge-dominated
galaxies and disk-dominated galaxies \protect\citep{bell04}.
Work on automated diagnostics of galaxy interactions
in GEMS is ongoing, and will be discussed in a future work.    }.
In what follows, we adopt by-eye classification bins of E/S0, Sa--Sd, 
irregular/compact (Irr), and peculiar/clearly interacting (Pec/Int).
Galaxies were classified on the basis of both central light concentration
and smoothness.  E/S0 galaxies were required to have dominant
spheroids and smooth light distributions.  Sa--Sd galaxies were
required to have prominent disks with signs of ongoing star formation.
Galaxies classified as irregular are chosen to be similar 
in morphology to low surface brightness and Magellanic irregular
galaxies in the local universe; their irregular light distributions
appear to the eye to result from stochastic variations
and bursts in their star formation history rather than 
from any interaction.  In contrast, 
the peculiar/clearly interacting designation 
requires that galaxies have signs of tidal features
and/or multiple nuclei, with morphologies
suggestive of a major galaxy merger or merger remnant\footnote{In 
many works galaxies that do not fit onto
the Hubble sequence are labeled irregular, where no 
attempt is made to characterize the driver
of the irregularities.  Because we wish to gain insight
into the role of interactions in driving the star 
formation activity of galaxies, we attempt, perhaps naively, to 
distinguish between galaxies with clear signs of interaction (Pec/Int) and
those without (Irr).}.
These classifications are discussed in detail
and examples of each class shown in \citet{wolfuv}. 

Obviously, these classifications are subjective and depend
on the depth of the imaging material.  In particular, comparison
with deeper GOODS data for a subset of 
290 galaxies in this redshift slice has shown that the 
fraction of faint clearly interacting galaxies in GEMS 
is slightly underestimated \citep{wolfuv}.
While the ideal solution would be to simply consider only the GOODS data,
number statistics is a key limitation: there are $\sim 80$ 
galaxies at $z \sim 0.7$ in the GOODS area that are detected 
at 24{\micron}, compared to the $\ga 400$ galaxies detected in 
the whole GEMS area.  We therefore choose to study galaxies 
from the whole GEMS area, adjusting the GEMS-only classifications
as a function of IR luminosity
to account for the underestimated faint interaction fraction.
We address this topic, and the 
scatter among the three different classifiers where relevant
in \S \ref{sec:res1} and Appendix \ref{app:goods}.

\subsection{Cross-correlation of the IR and optical data}

At this stage, we have simply correlated the 24{\micron}
imaging catalog with the COMBO-17 catalog.  The 24{\micron} footprint does not 
exactly match that of COMBO-17; of the 1727 COMBO-17 
galaxies in our redshift
slice, 1436 are covered by the 24{\micron} data.  
A subsample of 1306 of these are covered by the GEMS mosaic, 
and have full morphological, 24{\micron} and redshift information.
Sources are matched within 2\arcsec.  Roughly
96\% of 24{\micron} sources have only one match 
in the COMBO-17 catalog. The other 4\% have two matches in the COMBO-17
catalog; the closest positional match is 
taken in these cases\footnote{This
effect may lead to 1--2\% incompleteness if two sources with 
24{\micron} emission are mistakenly blended into one source}.
A 24{\micron} catalog constructed using optical positions as prior
constraints would address many of these limitations, but is 
challenging to implement at this stage; we defer a complete analysis of
this type to a later date.
In total, 442 galaxies with $0.65 \le z < 0.75$ 
are detected at 24{\micron} and have
a match in COMBO-17 (397 overlap with GEMS);
the rest of the COMBO-17 sources
without 24{\micron} detections are assigned 
$5\sigma$ upper limits of 83$\mu$Jy\footnote{Sources near
the map edges where the noise is non-uniform have been disregarded 
in this work.}.  
In appendix \ref{app:c17comp} we explore
the possibility that our sample selection 
may be biased against highly-obscured IR-luminous
galaxies. We find that $z \sim 0.7$ IR-luminous
galaxies should be bright enough to be detected and 
successfully classified by COMBO-17, suggesting that 
the sample selection and main conclusions of this paper 
should be robust to this potential source of incompleteness.

\section{Analysis} \label{sec:ana}

\subsection{Estimating total IR luminosity} \label{subsec:esttot}

Local IR-luminous galaxies show
a tight correlation between 
rest-frame 12--15{\micron} luminosity and total IR luminosity
\citep[e.g.,][]{spi95,cha01,rou01,papovich02}, with a scatter
of $\sim 0.15$ dex.  Following \citet{papovich02}, we 
use this correlation to construct total IR luminosity from the observed-frame
24{\micron} data (corresponding to rest-frame $\sim 14${\micron}).
We use the full range of \citet{dale01} model template spectra
to calculate the conversion from observed-frame 24{\micron} to 
8--1000{\micron} total IR luminosity\footnote{Total 
8--1000{\micron} IR luminosities are $\sim 0.3$ dex 
higher than the 42.5--122.5{\micron} luminosities defined by \citet{helou88},
with an obvious dust 
temperature dependence. }; 
we adopt the mean value as our estimate of 
IR luminosity and the RMS as our estimate of conversion uncertainty.
The mean correction factor corresponds approximately to  
$L_{\rm IR} \sim 10 \nu l_{\nu}$(24{\micron}) to within 20\% 
where $l_{\nu}$(24{\micron}) is the monochromatic luminosity
at observed-frame 24{\micron}; the template scatter
around this mean is larger, $\sim 0.3$\,dex.  Different
choices of template \citep[e.g.,][]{dev99}, or the adoption
of a luminosity-dependent conversion yield similar results.
The $5\sigma$ detection limit of 83$\mu$Jy corresponds 
to limiting IR luminosity of $6 \times 10^{10} L_{\sun}$ or 
$\sim 6 {\rm M_{\sun}\,yr^{-1}}$ in terms of an IR-derived SFR.

The conversion from the 24{\micron} 
photometry to $8-1000${\micron} IR luminosity 
should be accurate a factor
of two or better, which is adequate for our purposes. 
There are two types of error. First, the conversion 
depends on fitting templates to the galaxy SEDs. There is a total range of a
factor of two difference between the conversion that would be predicted by 
various templates for galaxies in the 
appropriate luminosity range of $\sim 10^{11} L_{\sun}$  
\citep{dev99,dale01,cha01,lagache03,lagache04}.
Second, there is a possibility that the infrared properties 
of typical high-luminosity galaxies differ 
at z $\sim$ 0.7 from those of the local galaxies used in developing the
templates.  The evidence available to us now suggests that 
the IR SEDs of $z \ga 0.7$ seem to be adequately
spanned by the IR SEDs of local galaxies:
the evolution of the MIR--radio correlation to $z \sim 1$
\citep[e.g.,]{elbaz02,appleton04}, 15--24{\micron} flux ratios
from ISO and Spitzer for galaxies at $z \sim 1$ \citep{elbaz04}, 
and mid-IR spectra from Spitzer of $z\sim 1$ ULIRGs (L.\ Yan, private
communication).  This issue will be significantly
clarified by future work from Spitzer and Herschel.

\subsection{Estimating star formation rates}

We estimate SFRs
using the combined UV and IR emission of the sample galaxies.
This SFR estimator accounts for both 
direct light from young stars from the UV and 
obscured light from the IR, giving a complete census of 
the bolometric luminosity of young stars in the 
galaxy \citep[e.g.,][]{fluxrat}.  The primary difficulties affecting this
SFR estimator are (1) the poorly-constrained 
fraction of IR light from dust heated by old stars, and (2)
geometry, in the sense that the IR light is radiated 
isotropically, whereas the UV light escapes along preferred directions
(such as out of the plane of a disk galaxy).  
Future works will be able to address point (1) through 
more detailed modeling of UV/optical/IR SEDs, 
and will be able to ascertain to which extent 
averaging over inclination angles effectively 
addresses point (2).  

In practice, we estimate the SFR $\psi$ using a calibration 
derived from the PEGASE 
stellar population models \citep[see][for a description of an earlier version of the model]{fioc97}, assuming a 100\,Myr-old stellar population 
with constant SFR and a \citet{kroupa01} IMF: 
\begin{equation}
\psi / ({\rm M_{\sun}\,yr^{-1}}) = 9.8 \times 10^{-11} \times
       (L_{\rm IR} + 2.2L_{\rm UV}),  \label{eqn:sfr}
\end{equation}
where $L_{\rm IR}$ is the total IR luminosity 
and $L_{\rm UV} = 1.5 \nu l_{\nu,2800}$ is a rough estimate
of the total integrated 1216{\AA}--3000{\AA} UV luminosity, derived using
the 2800{\AA} rest-frame luminosity $l_{\nu,2800}$ from COMBO-17.
The factor of 1.5 in the 2800{\AA}-to-total UV conversion accounts
for the UV spectral shape of a 100 Myr-old population with constant
SFR.  The SFR is derived assuming that $L_{\rm IR}$ reflects the bolometric
luminosity of young, completely obscured populations, and 
that $L_{\rm UV}$ reflects the contribution of unobscured stars, 
that must be multiplied by a factor of 2.2 to account for
light emitted longwards of 3000{\AA} and shortwards of 1216{\AA} by 
the unobscured young stars.  In practice, the SFRs are dominated
by the IR contribution for the range of SFRs explored
in this paper.
This SFR calibration is derived under identical assumptions as 
the UV and IR SFR calibrations presented by \citet{k98}.
Our SFR calibration is consistent with \citet{k98} to 
better than 30\% once different IMFs are accounted for. 
\citet{bellsfr} tested UV$+$IR-derived SFRs
against carefully extinction-corrected \ha and radio-derived SFRs for
a large sample of local star-forming galaxies, 
finding excellent agreement with 0.3 dex scatter and no offset. Accordingly,
we adopt a systematic error estimate of 0.3 dex (dominated
by 24{\micron}-to-total conversion error) and 0.4
dex random error (with contributions
from both 24{\micron}-to-total conversion and 
limitations in UV$+$IR-derived SFRs).
These error estimates are supported by comparisons
between ISO 15{\micron}-derived SFRs 
with carefully-constructed Balmer-line
SFRs for galaxies at $0 < z < 0.8$: there was no offset between 
the SFR scales and 0.4 dex 
random scatter for SFRs less than 250${\rm M_{\sun}\,yr^{-1}}$
\citep{flores04}.

\subsection{Estimating stellar mass}

To understand how the instantaneous SFR in a galaxy
compares with its mean past rate, it is necessary to 
estimate the existing stellar mass.  
Under the assumption of a universally-applicable 
stellar IMF there is a tight correlation 
between rest-frame optical color and stellar M/L, that is rather insensitive
to details of galaxy SFH, dust content and metallicity
\citep[e.g.,][]{bdj,bell03,kauffmann03}. Based on this correlation, we estimate stellar
mass $M_*$ using the rest-frame $B$ and $V$ band magnitudes:
\begin{equation}
\log_{10} M_* / M_{\sun} = -0.4(V - 4.82) + (-0.628 + 1.305 [B-V] -0.15),  \label{eqn:sm}
\end{equation}
where the $-0.15$ dex term converts the stellar masses to a 
\citet{kroupa01} IMF\footnote{More sophisticated stellar 
mass estimates, based on direct modeling of COMBO-17's 
observed 17-passband SED (Borch et al., in preparation), 
agree with these color-based
estimates with no systematic offset and $\la 40$\% random scatter.}.
Systematic and random errors in 
stellar masses are roughly 0.3 dex, owing 
primarily to the effects of dust, galaxy age and bursts of star 
formation \citep{bdj,bell03,fran03}.  
For stellar populations completely dominated
by $\la 1$\,Gyr-old stars, the luminosity at a given 
color is roughly a factor of three higher than for galaxies
with a wider range of stellar ages.  Because Eqn.\ \ref{eqn:sm}
assumes a wide range in stellar age, the stellar
masses of galaxies with recent very large 
bursts of star formation will be {\it overestimated}.

\section{Trends in IR-to-UV ratio with SFR}
             \label{sec:iruv}

\begin{figure}[tb]
\begin{center}
\epsfxsize=9cm
\epsfbox{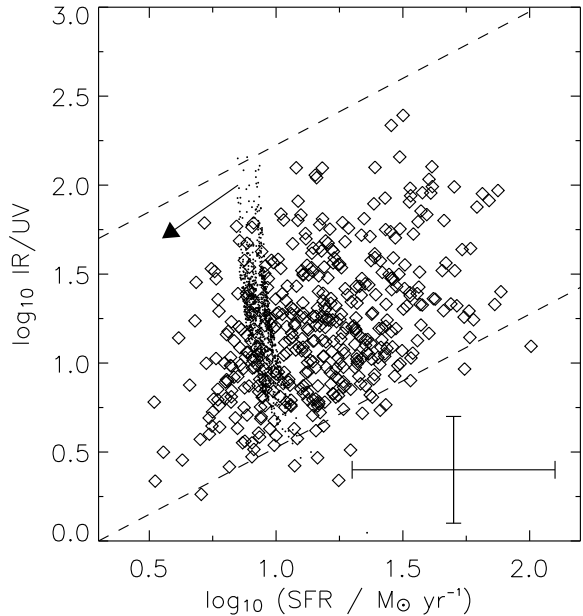}
\caption{\label{fig:iruv} IR/UV as a function 
of SFR.  Open diamonds show galaxies detected with MIPS at 24{\micron}, 
whereas points show 24{\micron} upper limits for galaxies not detected
by MIPS; 
these limits will move towards the lower left with a slope of 
unity as shown by the arrow, and are therefore are 
consistent with the trend observed for the detected galaxies.
The dashed lines outline the local relation as derived by \citet{adel00}.
The error bars show our uncertainty estimates: 0.3 dex in IR/UV
(dominated by 24{\micron}-to-total conversion
uncertainty) and a 0.4 dex SFR uncertainty.
 }
\end{center}
\end{figure}

The comparison of UV and IR luminosities shows that, both on a
galaxy-to-galaxy basis and globally, the SFR estimates 
discussed below are essentially IR-derived, with a minor 
$\la$ 30\% contribution from the UV. To assess the relative 
importance of UV and IR 
light in individual galaxies at $z \sim 0.7$, Fig.\ \ref{fig:iruv} 
shows IR/UV as a function of SFR, as derived from the IR and UV luminosities
using Eqn.\ \ref{eqn:sfr}.  
The range of IR/UV and the trend with total luminosity in the z
$\sim$ 0.7 sample is very similar to that observed in the local Universe 
\citep[e.g.,][]{wang96,adel00,bellsfr}, indicating little or no evolution
in the relative proportion of obscured star 
formation at least to $z \sim 0.7$\footnote{This is also
consistent with a sample of very luminous IR-luminous $z \sim 1$ starbursts
from a 15{\micron}-selected 
ISO sample \citep[as discussed by Adelberger \& Steidel]{aussel99}.}. 
This is illustrated in Fig.\ \ref{fig:iruv}, where
the locus of local star-forming galaxies is also shown\footnote{
This has been adapted from their Fig.\ 11a, 
where $\log_{10} {\rm SFR} \sim \log_{10}(L_{1600}+L_{\rm bol,dust}) - 9.7$,
accounting for our adoption of $H_0 = 70$\,km\,s$^{-1}$\,Mpc$^{-1}$.
The UV luminosity is estimated by $L_{\rm UV} \sim \nu f_{\nu,1600}$ 
(derived using PEGASE for a 100Myr-old stellar population 
with constant SFR; coincidentally, the UV spectral shape 
cancels out, leaving 1216{\AA}--3000{\AA} luminosity 
equal to monochromatic 1600{\AA} luminosity).
These estimates of SFR and IR/UV should be consistent with 
our estimates, to within the considerable systematic uncertainties. }. 
Galaxies with SFRs $\ga 6 M_{\sun}$ yr$^{-1}$ have 
IR/UV $\ga$ 10, indicating that uncorrected UV 
SFRs are gross underestimates of the true SFR 
for the most intensely star-forming systems.  

The cumulated total IR luminosity of all detected sources in 
this redshift slice is $7 \pm 3
\times 10^{13} L_{\sun}$, where the error bar is dominated 
by our assumed error in 24$\mu$m to total IR conversion\footnote{This
is the total luminosity in the volume-limited $0.65 \le z < 0.75$
sample, and so is in essence a luminosity density.  We choose
to express quantities in terms of total luminosities at
this time owing to concerns about cosmic variance.  This topic
is considered in more detail by Le Floc'h et al.\ (in preparation).}. 
If we assign detections at 83$\mu$Jy to all
the upper limits, we derive an upper limit to the total 
IR luminosity of twice the value from the 
detected galaxies. This result agrees with the 
number counts and our current IR luminosity function estimates, 
which indicate that about 
2/3 of the total background at 24$\mu$m is resolved at our detection limit
\citep[Le Floc'h et al., in preparation]{papovich04}.  
We conclude that the total IR 
luminosity in this redshift slice is $\sim 10^{14} L_{\sun}$.  
In comparison, the total UV luminosity in this redshift slice is 
$1.5 \pm 0.3 \times 10^{13} L_{\sun}$ when the luminosity 
function is extrapolated to zero luminosity \citep[using the 2800{\AA}
luminosity function determined for this redshift 
slice by][]{wolfuv}\footnote{It is worth noting that there is a
difference in how we define luminosity densities, compared
with \citet{wolf03} and \citet{wolfuv}.  
These papers express luminosity densities
in terms of monochromatic solar luminosities at a given rest-frame
wavelength, whereas we 
choose to present $L_{\rm UV}$ in terms of bolometric solar luminosities.}. 
Thus, at $z \sim 0.7$, the ratio of IR to UV luminosity density is 
$\sim 7_{-3}^{+4}$.

\section{Exploring the relationship between star formation rate and morphology} 
\label{sec:res1}

\begin{figure*}[tb]
\begin{center}
\epsfxsize=15cm
\hspace{0.3cm}\epsfbox{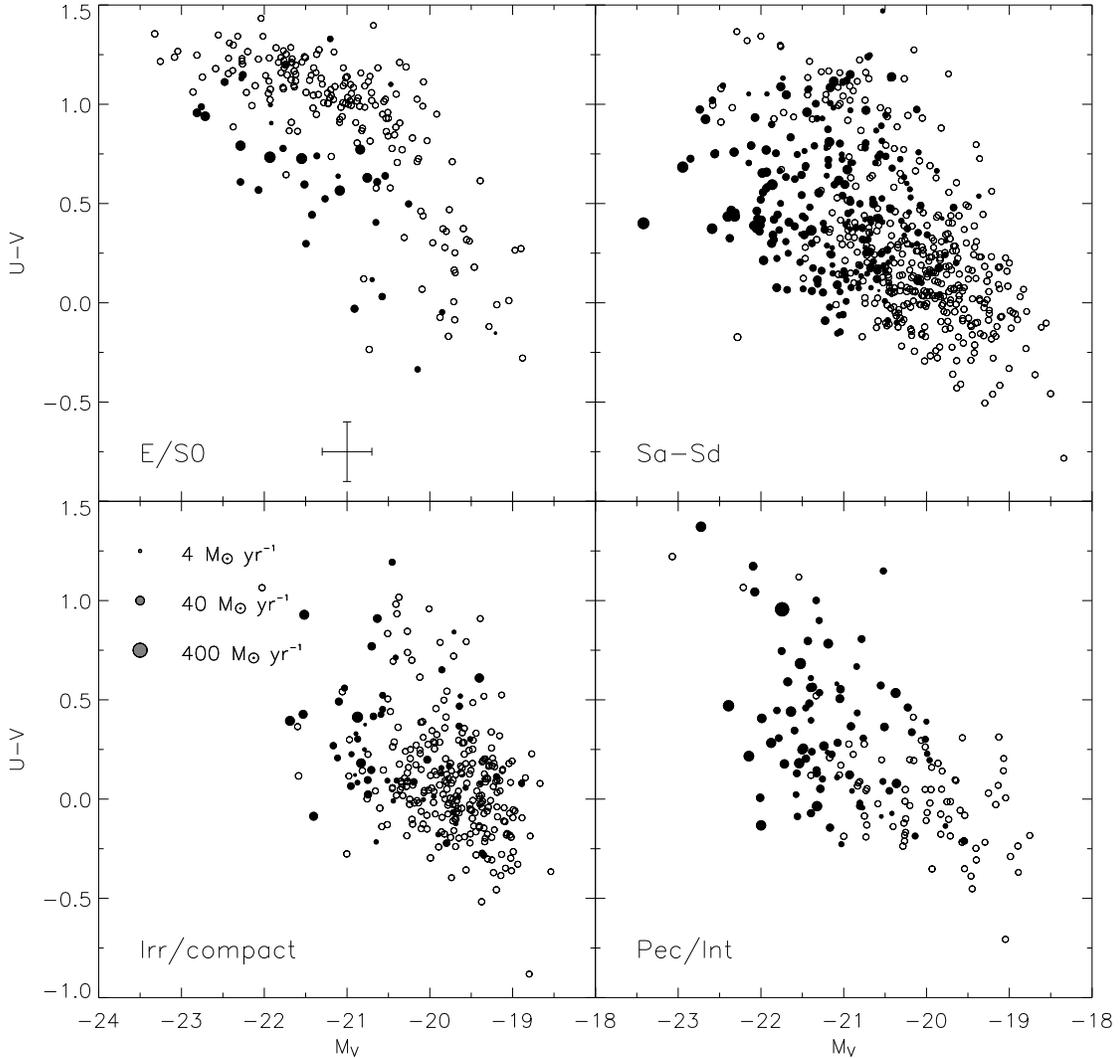}
\caption{\label{fig:cmr} The rest-frame $U - V$ color of 1436
optically-selected 
$0.65 \le z < 0.75$ galaxies from COMBO-17/GEMS, as a function of 
visually-classified rest-frame $V$-band galaxy morphology (classifications
are adopted at random from three classifiers).  Symbol
size scales with the logarithm of the estimated SFR: 
filled circles denote galaxies with 
MIPS 24$\micron$ detections (therefore the symbol size denotes
SFR estimated using IR$+$UV) whereas the open circles show upper
limits on the SFR, derived using the observed UV luminosity and
the upper limits on 24{\micron} flux.
}
\end{center}
\end{figure*}

In this section, we 
explore which types of galaxies account for the bulk of star formation in 
the $z \sim 0.7$ universe, when the universe was only half of its present age.

\subsection{The relationship between optical properties and star formation}

In Fig.\ \ref{fig:cmr}, we show the rest-frame optical
$U-V$ color and $V$-band absolute magnitude of 
all galaxies in the $0.65 \le z < 0.75$ slice, 
split by visual morphological type.
Filled symbols denote galaxies detected at 24{\micron}
while open symbols
denote galaxies for which only 24{\micron} upper limits (83$\mu$Jy) 
were available.  In both cases, symbol size scales with the logarithm 
of SFR or its upper limit.  

Among visually-classified E/S0 galaxies, those that are 
blue in the optical are also clearly detected
at 24{\micron}.  Ignoring for the moment active nuclei
(see \S \ref{sec:disc}), we attribute this 24{\micron} flux to
rapid star formation.  
In contrast, there are very few galaxies on the red sequence with
24{\micron} detections.  Thus, red E/S0 galaxies are IR-faint, which 
suggests that their red optical colors are the result of dominant
old stellar populations, not dust-obscured star formation.

Blue visually-classified spiral galaxies (Sa--Sd)
are frequently detected in the IR.  Many optically-red
spiral galaxies are IR-faint, and are thus red because of 
their ancient stellar populations.  
Yet, a rather higher fraction of red 
spirals are detected at 24{\micron} compared to red E/S0 galaxies,
especially at fainter optical luminosities,
indicating a more important role for dust-reddened
spiral galaxies.  Morphologically-classified irregular
galaxies are typically somewhat less luminous, and owing to their
lower SFRs are less frequently 
detected by MIPS; however, they
are qualitatively similar to low-luminosity spirals.

The vast majority of visually-classified peculiar
and clearly interacting galaxies are
detected at 24{\micron}; indeed, the few non-detections are almost
all relatively faint and are not expected to be detected at 24{\micron}
for plausible UV/IR ratios.  Interestingly, most red Pec/Int galaxies
are strong 24{\micron} sources (including the brightest galaxy at 
24{\micron} in the
$0.65 \le z < 0.75$ slice, which is an interacting galaxy 
on the blue edge of the red sequence).  Therefore, we conclude that
{\it i)} almost all of our visually-classified Int/Pec galaxies are
intensely star-forming at the epoch of observation, and {\it ii)} 
that the optical colors of Int/Pec galaxies largely 
reflect their dust content, rather
than their SFH.

In summary, the optical color of morphologically-undisturbed
galaxies (E/S0 and Sa--Sd galaxies) reflects SFH in a broad sense:
most red early-type galaxies, and a significant fraction of red 
spiral galaxies, are non-star-forming; whereas
the blue early and late-type galaxies have 
important amounts of current star formation as probed by the IR.
In contrast, clearly interacting and peculiar galaxies have
strong IR-luminous star formation irrespective of their optical properties
(as is expected of very dusty star-forming systems).  

\subsection{The contribution of different galaxy types to 
the cosmic SFR}

An important question is the relative 
contribution of galaxies with different morphological types to the 
total SFR density at $z \sim 0.7$.  The cosmic-averaged
SFR has dropped by a factor of three since 
$z \sim 0.7$ as measured in the 
IR \citep[e.g.,][]{flores99,pozzi04}, radio
\citep[e.g.,][]{haarsma00}, H$\alpha$ \citep[e.g.,][]{tresse02,perez03},
or UV \citep[e.g.,][]{lilly96,schiminovich}.
Thus, an understanding of the relative contribution 
of different galaxy types to the cosmic SFR at $z \sim 0.7$
will constrain the types of galaxy that need 
to `switch off' their star formation activity between $z \sim 0.7$
and the present day, 
giving insight into the physical mechanisms
driving this evolution.  

\begin{figure}[tb]
\begin{center}
\epsfxsize=9cm
\epsfbox{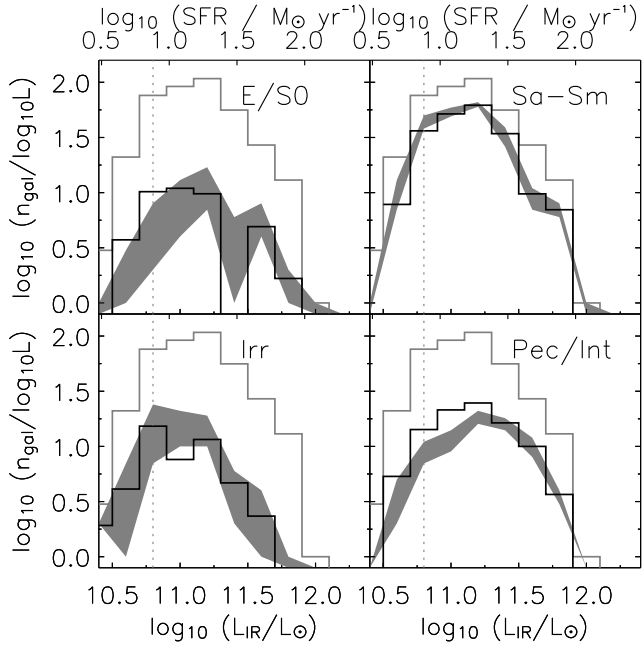}
\caption{\label{fig:lf} The estimated 8--1000{\micron} 
luminosity function, split by morphological type for 
397 galaxies at $0.65 \le z < 0.75$ 
galaxies.  Only galaxies detected at 24{\micron} are shown, 
and no attempt to extrapolate to lower IR luminosities has been made;
the sample is grossly incomplete below $\sim 6\times10^{10} L_{\sun}$
as denoted by the grey dotted line.
In each panel, the grey solid histogram shows the 
total IR luminosity function.  The shaded area shows
the IR luminosity function split by galaxy type 
using GEMS-derived galaxy classifications, where the 
extent of the shaded area shows explicitly the 
differences in IR luminosity function given by the 
three different classifiers.  The black histogram
shows the IR luminosity function, averaged over the 
three different classifiers and corrected to reproduce
the increased fraction of clearly interacting galaxies
seen in GOODS-depth data, as described in Appendix \ref{app:goods}.
 }
\end{center}
\end{figure}

In Fig.\ \ref{fig:lf}, we show the distribution of estimated IR luminosity
as a function of galaxy type for 24{\micron}-detected  
galaxies at $z \sim 0.7$; the top axes show the equivalent
SFRs.  We choose not to normalize the luminosity function to 
the observed volume at this stage owing to concerns about cosmic
variance \citep[e.g.,][]{somer04}; a thorough analysis of 
evolution in the IR luminosity function is presented
in a companion paper by Le Floc'h et al. (in preparation).
We do not extrapolate the IR luminosity function 
below the 24{\micron} detection limit at $\sim 6\times10^{10} L_{\sun}$.

The IR luminosity density for 
$10.7 \la \log_{10} L_{\rm IR}/L_{\sun} \la 11.5$ is
dominated by spiral galaxies, with an important contribution from clearly
interacting galaxies.  Morphologically-classified E/S0 galaxies and 
irregulars contribute, but play lesser roles.    Clearly
interacting galaxies make an increasingly important contribution 
at $\log_{10} L_{\rm IR}/L_{\sun} \ga 11.3$. 
This behavior resembles that in the local universe,
where almost all galaxies with 
$\log_{10} L_{\rm IR}/L_{\sun} \ga 11.8$ show clear signs 
of interaction \citep{sanders96}.  Integrating over the whole
distribution of galaxies detected at 24{\micron},
we find that the fractional contributions of spiral, 
E/S0, Pec/Int and irregular galaxies to 
the total integrated IR luminosity density are 0.51 : 0.10 : 0.28 : 0.11.

To probe the uncertainties in the relative contributions to 
the SFR density from different galaxy types, we explore two
limiting cases: (1) all 24{\micron} non-detections
have no IR flux (i.e., their SFR is measured purely from the UV); or
(2) all 24{\micron} non-detections emit at the detection upper
limit at this wavelength of 83$\mu$Jy.  
In the first case, the split of SFR is (Spiral : E/S0 : Pec/Int : Irr)
0.51 : 0.10 : 0.26 : 0.13, very similar to the split in terms of 
observed IR flux (because the UV is a relatively small 
contribution to the total SFR density).  In the second case, where
limits are all taken to be marginal detections, the split
is 0.47 : 0.15 : 0.18 : 0.20.  Because of the preponderance of 
E/S0 and Irr non-detections, the relative importance of these 
classes is increased (arguably, rather artificially).  
In neither of these cases are
the contributions from optically-faint galaxies
(i.e., those galaxies fainter than $M_V \ga -19$)
accounted for.  Because the integral over a luminosity
function is dominated by $\sim L*$ galaxies, our
limitation to relatively bright sources does not
prevent us from qualitatively constraining the 
contribution of different galaxy types. Nonetheless,
it is clear that the detailed form of the distribution of
SFRs to lower limits will importantly affect the quantitative
contributions of the different galaxy types, and is arguably
the largest source of systematic uncertainty in this paper.

To summarize, the SFR density at $z \sim 0.7$ 
is dominated by morphologically-normal spiral, E/S0, and irregular
galaxies ($\ga 70$\%).  Clearly interacting galaxies account 
for $\la$ 30\%. Our result is based 
on a combination of UV and IR luminosities, 
but is dominated by IR-luminous objects. 
\cite{wolfuv} have examined the UV-luminous objects alone and
also find that most of the rest-frame UV flux density at $z \sim 0.7$ is in
morphologically-undisturbed galaxies.

These results have important implications 
for the physical mechanisms which 
drive the factor-of-three decline in cosmic-averaged comoving SFR density
from $z \sim 0.7$ to the present day.  
Because clearly interacting galaxies ---
with morphologies suggestive of major galaxy mergers or merger
remnants\footnote{It 
is worth noting that relatively advanced
merger remnants are also classified as clear interactions, until
their tidal features have dissipated.  Owing to the 
long dynamical times at large galactic radii, we expect that
major mergers are classified as clearly interacting well
after the major final burst of star formation has ceased.} --- contribute
much less than half of the $z \sim 0.7$ SFR density, the 
drastic decline in cosmic SFR is only {\it partially} driven by 
the decreasing frequency of major galaxy mergers at the present 
day \citep[e.g.,][]{lefevre00,patton02}.  Instead, the dominant
driver is a strong decrease in SFR in morphologically-undisturbed spiral
galaxies.  It is
possible that the
decreasing star-formation activity of spiral galaxies
is driven primarily by gas supply and consumption in a purely
quiescent mode of star formation.  However, it is also 
possible that even in spiral galaxies much star-formation 
activity is triggered by relatively minor tidal interactions, that 
function to trigger internal instabilities such as spiral arms or 
bars, enhancing their SFR.  Our observations cannot 
sensitively test these competing hypotheses at this stage: deeper
redshift surveys and exploration of their gas content through, e.g.,
CO mass determination from the {\it Atacama Large Millimeter Array}
will start to allow us to test the relative importance of these
physical processes.  

\section{Exploring the demise of star formation in massive galaxies}  
   \label{sec:massive}

\begin{figure*}[tb]
\begin{center}
\epsfxsize=15cm
\hspace{0.3cm}\epsfbox{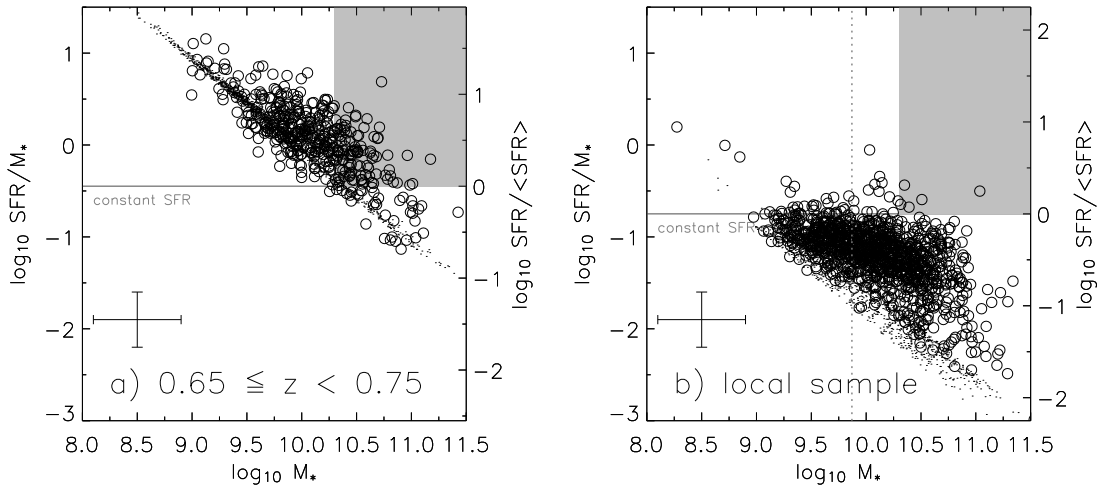}
\caption{\label{fig:br}  
Specific SFR as a function of  
galaxy stellar mass, in solar units.  {\it a)} 
SFR per Gyr per unit stellar mass as a function of stellar
mass for the $z \sim 0.7$ galaxy sample.  
Plotted on the right-hand axis is the birthrate parameter
$b = {\rm SFR / \langle SFR \rangle}$.  Galaxies detected
at 24{\micron} are shown with open circles; non-detections
have upper limits shown by points.  
{\it b)} SFR per Gyr per unit stellar mass as a function of stellar
mass for the local comparison sample.  Again, the birthrate
parameter is plotted on the right-hand axis for comparison 
(the offset between panels {\it a} and {\it b} is because 
of the difference in the age of the universe at the two redshift
intervals of interest).
IR detections are shown 
as open circles; non-detections are denoted by points.  The dotted line 
shows the minimum mass for which the sample is complete.  In both panels,
the solid grey line shows a birthrate of unity; i.e., a constant SFR through
all of cosmic history.  The grey box shows the position of 
intermediate- and high-mass galaxies
with ongoing bursts of significant star formation ($b > 1$).  Both datasets
sample $\sim 1.2 \times 10^5$ Mpc$^3$, therefore direct comparison of both 
plots is fair, remembering the influence of the detection limits.
}
\end{center}
\end{figure*}

We turn now to explore the relationship between SFR and stellar mass
at $z \sim 0.7$, comparing it to $z \sim 0$.
We choose to explore the `specific SFR'; i.e., the SFR
per unit stellar mass.
This is closely related to 
the more model-dependent 
birthrate parameter, $b = {\rm SFR} / \langle {\rm SFR} \rangle$,
where $\langle {\rm SFR} \rangle$ is the past-averaged SFR.  The
past-averaged SFR is estimated by dividing the stellar mass by the
elapsed time since star formation began in a given galaxy 
\citep[see, e.g.,][]{ke94}, adjusting for the mass lost by 
the stellar population as it ages.  Roughly 50\% of initially-formed
stellar mass is lost within one billion years for a \citet{kroupa01}
IMF; therefore $b \simeq {\rm SFR} / (2M_*/t_{sf})$.  In 
estimating $t_{sf}$, we simplistically assume the epoch of first 
star formation is at $z_f = 4$ for all galaxies; using a present-day age of 
the Universe of 13.5 Gyr, this corresponds
to $t_{sf} \sim 12$\,Gyr at $z \sim 0$ and 
$t_{sf} \sim 6$\,Gyr at $z \sim 0.7$.
Despite the extra model dependence introduced by 
having to assume a timescale when deriving $b$, 
birthrate has the attractive feature that it compares the 
observed SFR with the SFR required to build up the existing stellar
mass within a Hubble time (at that epoch).  

In panel {\it a)} of Fig.\ \ref{fig:br}, 
we show specific SFR as a function of stellar mass
for the 1436 $0.65 \le z < 0.75$ galaxies in our sample.  
Owing to the detection limit of
the 24{\micron} observations, it is impossible to detect 
galaxies forming stars at anything less than a constant rate 
(SFR/$M_* \la 0.4$, or equivalently $b \la 1$)
for stellar masses less than $\sim 2 \times 10^{10} M_{\sun}$.  
Yet, despite this rather stringent limitation, large numbers of galaxies
have been detected at 24{\micron}, with typical 
specific SFRs of $\sim$1/Gyr (i.e., if the 
current SFR were to continue at its present rate
the stellar mass of the galaxy would double within 
2 Gyr, remembering that 50\% of the mass
initially in stars is returned to the 
gas phase by stellar winds and supernovae). 

Given the existence of an important population of 
intensely star-forming galaxies at $z \sim 0.7$, it
is instructive to explore how this compares with 
the properties of star-forming galaxies in the local
universe.  We have therefore constructed a local `control'
sample of galaxies, which we show in panel {\it b)} of Fig.\ \ref{fig:br}.  
We use the NASA/IPAC Extragalactic 
Database (NED) to select a pseudo volume-limited sample 
in the range $1500 \le cz/{\rm km\,s^{-1}} \le 3000$
with 2MASS $K$-band magnitude cut $K < 12$,
and galactic latitude $b > 30\deg$.  This sample
is complete for galaxies above $M_K < -19.6$.
Where they were defined, 
IR fluxes were taken also from NED; total 8--1000{\micron} 
IR fluxes were estimated following Appendix A of \citet{bellsfr}, and 
are accurate to $\sim 30$\%.  Upper limits on IR flux were roughly estimated
assuming total IR fluxes just at the IRAS detection limit of 
$\sim 3 \times 10^{-11} {\rm ergs\,cm^{-2}\,s^{-1}}$ (the conclusions
do not depend on the choice of this limiting flux).
A legitimate and unavoidable concern is the inhomogeneous completeness 
properties of NED redshifts.  For the purposes of volume estimation,
we estimate an overall NED redshift completeness from the
number of $K < 12$, $b > 30\deg$ galaxies with redshifts divided
by the number of $K < 12$, $b > 30\deg$ galaxies both with and without
redshifts: 14350/23658 (or 61\%).  Thus, the estimated total volume 
probed is $71500 \times 0.61 $Mpc$^3$.  
The total sample is 2177 galaxies, 
1089 with IR detections and 1088 with only IR upper limits.
In fact, because of large-scale overdensities in this 
volume, there are $\sim 3$ times more galaxies in this
volume than one would expect on the basis of, e.g., the stellar
mass functions or $K$-band luminosity functions of \citet{cole01} or
\citet{bell03}, taking into account all the relevant selection limits.  
To reflect this overdensity, 
we assign an `effective volume' to the sample of 
$\sim 1.2 \times 10^5 $Mpc$^3$; this `effective volume' matches 
the actual volume in the $0.65 \le z < 0.75$ slice.  
Accordingly, we use this control sample
unaltered, with the rationale that in 
the two panels of Fig.\ \ref{fig:br} we are 
comparing galaxies from the same effective volumes at both 
cosmic epochs.  None of our conclusions are affected by the detailed choice
of volume or, indeed, comparison sample.  Different 
NED-derived samples can be chosen, or a sample of GEMS/MIPS galaxies
with redshifts $0.1 \le z < 0.2$ can be used as a control, 
with unchanged results but a reduced number of comparison galaxies.

Stellar masses and SFRs for the local comparison sample 
were estimated assuming a Kroupa IMF.
Stellar masses were estimated directly from the $K$-band absolute 
magnitudes (where Hubble-flow distances are used, for simplicity)
assuming a single $K$-band stellar M/L of 0.6 in solar
units, appropriate for a Kroupa IMF 
\citep[following][]{bell03,cole01}.  SFRs were estimated
from total IR flux following equation 4 of \citet{bellsfr} adjusted
to a Kroupa IMF, and are not corrected for any (typically modest; $\la 30$\%) 
contribution to the dust
heating from old stellar populations 
(like the SFR estimates constructed using the 24{\micron}
data).  The stellar masses
and SFR estimates are accurate to $\sim 0.3$ dex, assuming 
a universally-applicable stellar IMF, and are directly comparable 
with stellar masses and SFRs for the $z \sim 0.7$ redshift slice.

\begin{figure}[tb]
\begin{center}
\epsfxsize=9cm
\epsfbox{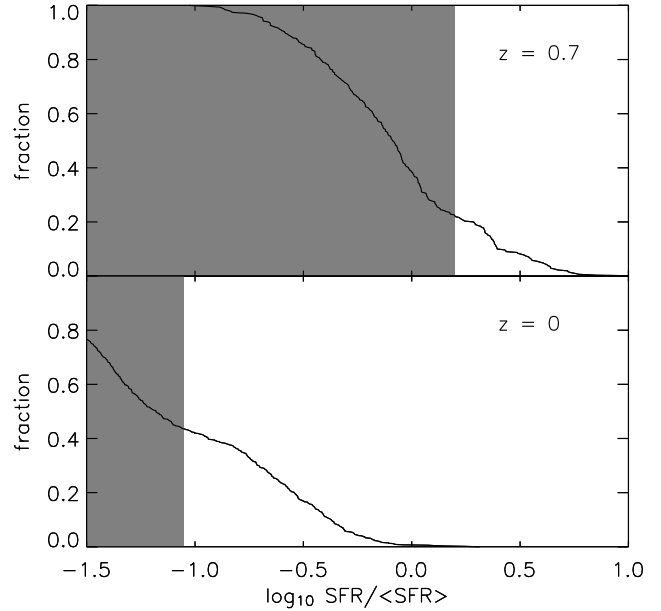}
\caption{\label{fig:brhist}  
The cumulative distribution of the birthrates of massive galaxies with 
$M_* > 2 \times 10^{10} M_{\sun}$.  The grey shaded region denotes
the area over which upper limits on birthrates contribute
to the distribution (i.e., the shape of the cumulative distribution in 
the grey shaded area is strongly biased towards high birthrates).
 }
\end{center}
\end{figure}

We compare the distribution of specific SFRs and stellar masses in 
the $z \sim 0.7$ sample and the local control sample in Fig.\ \ref{fig:br};
the cumulative histogram of birthrates is given in Fig.\ \ref{fig:brhist}.
Both datasets effectively sample $\sim 1.2 \times 10^5 $ Mpc$^3$, 
therefore in the regions where both datasets are complete it is 
fair to compare the relative distributions of datapoints.
The local sample is characterized by a broad distribution of 
star-forming galaxies with specific
SFRs between 0.03 and 0.2 ($0.1 \la b \la 1$; i.e., 
slowly declining SFR through cosmic history).
Furthermore, where the detection
limits permit, one observes a sizeable population of galaxies with 
very low specific SFRs $\la 0.01$, indicative of a largely non-star-forming
subpopulation of (primarily early-type) galaxies.  There are
very few galaxies with bursts of star formation that are
adding significantly to their existing stellar mass
(i.e., birthrates $b \ga 1$).  In particular, at 
stellar masses in excess of $2 \times 10^{10} M_{\sun}$, where
the sample is essentially complete, 5/699 galaxies ($<1$\%) 
of galaxies have $b \ge 1$ (the grey shaded region)\footnote{Galaxies
with $M^* > 2 \times 10^{10} M_{\sun}$ contain roughly 70\%
of present-day stellar mass \citep{bell03}.}.  
While there are significant
systematic uncertainties in both SFR and stellar mass, 
it is impossible to accommodate a large population of 
starbursting intermediate- and high-mass galaxies in the local universe.
This result is insensitive to the adoption of a variety
of plausible prescriptions for SFR and stellar masses, 
and indeed the rarity of galaxies with $b \ge 1$ at
$M^* > 2 \times 10^{10} M_{\sun}$ is apparent from 
Fig.\ 24 of \citet{brinchmann04}, who analyse 
accurate emission line-derived SFRs from the Sloan Digital 
Sky Survey.

In stark contrast, turning to the $z \sim 0.7$ sample, there are
a large number of massive galaxies with intense star formation.
Quantitatively, at stellar masses in excess of $2 \times 10^{10} M_{\sun}$,
120/311 (39\%) of galaxies are forming stars at a rate
faster than their past-averaged rate ($b \ge 1$; the 
grey shaded region in Fig.\ \ref{fig:br})\footnote{If
starbursts were defined by high specific SFRs instead,
the difference between $z \sim 0.7$ and the present day would
be even more striking.}.
While there 
are always significant systematic uncertainties, 
including estimation of total IR luminosity, SFRs and 
stellar masses, these differences are large enough 
to be robust to systematic uncertainties of greater than a factor of
three\footnote{It is likely that our neglect of heating 
of dust by old stellar populations affects galaxies
with low $b$ values preferentially, artificially 
inflating their estimates of SFR and therefore $b$.
From inspection of Fig.\ \ref{fig:br} 
this is likely to affect the most massive galaxies
the most severely, and will affect the low
redshift sample more acutely than the $z \sim 0.7$
sample.  Thus, if the heating of dust by old stars
were accounted for by detailed SED modeling, it is 
likely that the conclusions would remain unchanged, or further strengthen.}.  
In addition, where possible we have minimized the 
opportunities for systematic error by using IR-derived
SFRs in both cases, and equivalent SFR and 
stellar mass calibrations.  Indeed, adoption of 
the $0.1 \le z < 0.2$ GEMS/MIPS sample as the `local' 
control sample yields only 1 galaxy with $b \ge 1$ out of 
41 galaxies with $M_* \ge 2 \times 10^{10} M_{\sun}$.
In this case, identical methodologies have been used
for the $0.1 \le z < 0.2$ and $0.65 \le z < 0.75$ samples.

Recalling the conclusions reached in 
\S \ref{sec:res1}, it is worth briefly commenting on the 
visual morphologies of galaxies with $M_* \ge 2 \times 10^{10} M_{\sun}$.
70/124 massive spiral (Sa-Sd) galaxies, 4/5 
massive irregular galaxies, and 18/23 Pec/Int galaxies
have $b \ge 1$, i.e. are undergoing significant bursts
of star formation.  In contrast, only 20/138 massive E/S0 
galaxies are forming stars intensely\footnote{The total is 
290 galaxies, which is lower than the COMBO-17$+$MIPS number 
of 311 galaxies owing to the smaller areal coverage of GEMS.}.  Therefore, 
roughly 1/2 of spiral, irregular and clearly interacting galaxies
are undergoing starbursts at $z \sim 0.7$.  Furthermore, 
the starbursting population 
is dominated by spiral galaxies rather than Pec/Int galaxies.
This reinforces the conclusions of \S \ref{sec:res1};
even amongst the most intensely star-forming $z \sim 0.7$
galaxies, the vast majority are not forming stars 
intensely because of a recent major merger.

Taken together, it is fair to conclude that 
while a large fraction of intermediate- and high-mass galaxies are 
forming stars intensely 7\,Gyr ago, almost 
none are at the present epoch.  Furthermore, 
these intensely star-forming galaxies are primarily
morphologically undisturbed, indicating that major mergers
are not responsible for the bulk of this intense star
formation.  Instead, physical properties that do not strongly
affect galaxy morphology --- for example, a larger abundance of gas 
at earlier times or weak interactions with small satellite galaxies --- 
appear to be responsible for the high SFRs.

\section{Discussion} \label{sec:disc}

\subsection{Comparison with previous results}

Our results confirm and extend the previously-observed rapid 
evolution in the number of IR-luminous galaxies at 
intermediate and high redshift \citep[e.g.,][]{elbaz99,cha01,papovich04},
and with the drastic decline in the space density of 
intensely star-forming
galaxies as probed with a variety of different SFR
indicators \citep[e.g.,][]{cowie96,wolf03,bauer05}.
Furthermore, we have been able to estimate the stellar
masses of these intensely star-forming systems, finding 
that almost all of the intensely star-forming galaxies with 
$L_{\rm IR} \ga 10^{11} L_{\sun}$ (luminous infrared galaxies; LIRGs) seen at
intermediate redshift have stellar masses in excess
of $\sim 10^{10} M_{\sun}$ (Fig.\ \ref{fig:br}).
This is in agreement with, e.g., \citet{fran03} and 
\citet{zheng04}, who
used optical and NIR data to estimate the stellar masses of 
ISOCAM 15{\micron}-selected LIRGs, finding stellar masses in excess of 
$\ga 10^{10} M_{\sun}$, and \citet{lefloch04}, who 
explore the properties of 26 $1 \la z \la 2.5$ IR-luminous
sources with Spitzer, finding significant stellar masses
for their sample of LIRGs.

Our conclusion that a decline in major interaction rate
is not the primary cause of the decline in cosmic SFR since $z \sim 0.7$
also confirms and extends previous observations.  
Follow-up observations of fields targeted by 
ISO tentatively suggested that the rapidly-evolving IR-luminous population may
not be 
composed exclusively of strongly-interacting galaxies. \citet{flores99}
found that only 6/16 galaxies in a sample with 
ISO and HST data had strongly disturbed 
morphologies, while the rest were disk-dominated, 
E/S0, or unresolved.  \citet{zheng04} extended their work 
to a sample of 36 ISO-selected LIRGs at z $\le$ 1.2,
finding that 17\% of their sample were obvious mergers, 
and placing an upper limit of 58\% on 
the merger fraction. Our study places these 
indications on firm ground, pushing to fainter 
limits of $\sim 6 \times 10^{10} L_{\sun}$ 
with an order of magnitude larger sample of 
galaxies.

\subsection{How much dust can be heated by AGN at $z \sim 0.7$?}

\begin{figure}[tb]
\begin{center}
\epsfxsize=9cm
\epsfbox{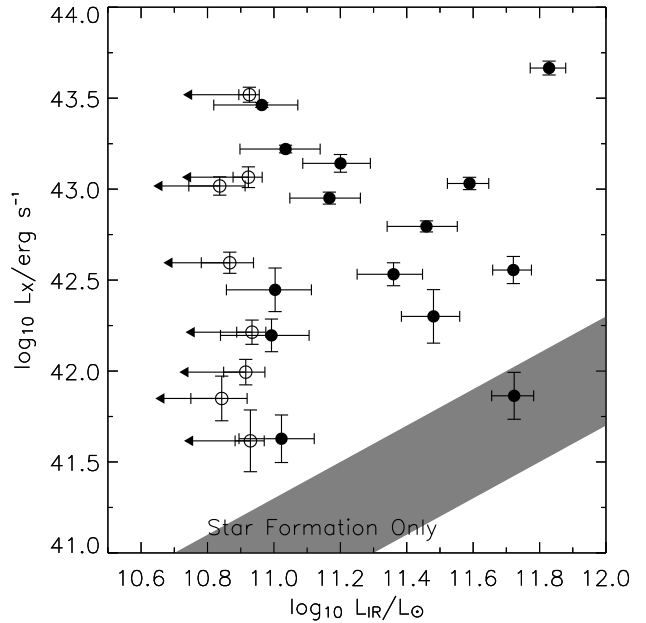}
\caption{\label{fig:x24}  
A comparison of 0.5--8 keV observed frame Chandra X-ray luminosities 
with 24{\micron}-derived
total IR luminosities for the 22 $0.65 \le z < 0.75$
Chandra-selected galaxies with optical 
detections.  Filled circles denote MIPS detections,
whereas open circles with limits denote MIPS upper limits only.  
The IR luminosity errors include uncertainty in conversion to total IR.
The shaded area shows the expected locus of star-forming
galaxies.  }
\vspace{-0.5cm}
\end{center}
\end{figure}

It is worth discussing the contribution 
of active galactic nuclei (AGN) to the 24{\micron} 
total luminosity density.  To aid in identifying AGN, 
we use the Chandra data for the CDFS to select
AGN candidates.  The Chandra
footprint on the CDFS \citep{alex03} is considerably smaller than the 
GEMS and MIPS coverage of the field, therefore in this section
we consider only galaxies falling completely within the 
Chandra coverage.  This area contains 3122 galaxies with COMBO-17
photometric redshifts; the small subsample classified
by COMBO-17 as galaxies with $0.65 \le z < 0.75$ and detected
by Chandra comprises 22 galaxies; 0.5--8 keV source fluxes
and errors are taken from \citet{alex03}.

While highly-luminous X-ray sources are invariably AGN, 
at fainter limits a contribution from star-forming galaxies 
is expected.  To test this possibility,
we compare 24{\micron}-derived IR luminosities and X-ray 
luminosities for these 22 $z \sim 0.7$ galaxies 
in Fig.\ \ref{fig:x24}.  The shaded area shows the expected locus
of star-forming galaxies, derived from the empirical relationship
between X-ray and $k$-corrected 1.4GHz 
radio luminosities \citep[Fig.\ 1 of][]{cohen03}
coupled with the radio SFR calibration of \citet{bell03}, accounting
for the 0.3 dex empirical scatter in the correlation\footnote{This
locus does not account for the mild X-ray $k$-corrections in
the observed frame 0.5--8 keV X-ray luminosity, 
but was defined for galaxies with $0.4 \la z \la 1.3$ with a median
redshift of 0.76 and so will be accurate
to $\sim 0.1$\,dex \citep{cohen03}.  It is worth noting that the 
use of the SFR calibration of \citet{cohen03} results in 
a similar locus to within 0.1 dex.}.
It is clear that only one galaxy
has X-ray emission that is consistent with a star formation origin;
the others have excess X-ray emission indicative of an AGN contribution.

Considering the 21  
AGN-dominated X-ray sources, we find a total IR luminosity between 
$3.4 \times 10^{12} L_{\sun}$ and $4 \times 10^{12} L_{\sun}$, where
in the first instance we ignore upper limits and in the second instance
we take upper limits to represent marginal detections.  The total 
IR luminosity in {\it all} detected sources in the Chandra footprint
is $2.6 \times 10^{13} L_{\sun}$, increasing to $5.4 \times 10^{13} L_{\sun}$
if all non-detections are taken as marginal detections at the 83$\mu$Jy
limit.  Accordingly,
we estimate that $\la 15$\% of the total IR $0.65 \le z < 0.75$
luminosity density is in sources with significant AGN emission.
Clearly, the contribution to the IR luminosity density of AGN-heated 
dust may be substantially lower, as the IR luminosity in these galaxies
may come from star formation.  On the other hand, 
we have not accounted
for any contribution of AGN to the IR luminosities of galaxies
not detected by {\it Chandra}; bearing in mind that the 
Chandra observations reach deep enough to start probing
star-formation-powered X-rays from galaxies with $L_{\rm IR} \sim
10^{11} L_{\sun}$, one hopes that the AGN in non-AGN dominated
galaxies will be weak and will contribute relatively little 
to the total IR.  
This conclusion is consistent with 
recent ISO and Spitzer results, finding 10--20\%
contributions at 15{\micron} and 24{\micron} respectively
\citep{fadda02,fran04}, and with \cite{silva04} who model
the AGN contribution to the IR background, suggesting that 10--20\% 
of the IR background at mid-IR wavelengths is from AGN and 
their host galaxies (the AGN alone account for 5\% of the IR 
background in this model).

\subsection{Comparison with Theoretical Expectations} \label{sec:model}

Gas-dynamical processes, such as star formation or feedback,
are currently very challenging to include in models
of galaxy formation and evolution owing largely to
important resolution limitations.  As a result of 
the complexity of these processes, it has been impossible
to converge on a robust description of star formation;
instead, a wide range of 
empirically-motivated prescriptions 
on kpc or larger scales have been adopted
\citep[e.g.,][]{somerville99,springel03a,barnes04}.  Current 
models tend to reproduce the cosmic SFH reasonably
well \citep[e.g.,][]{cole00,somerville01,springel03b}.  
A few models have started to incorporate detailed dust
prescriptions, and can model the evolution of 
IR-bright galaxies \citep[e.g.,][]{granato00,balland03}.
While the progress is very encouraging, detailed observations
such as the 24{\micron} number counts are rather poorly
reproduced by the models \citep{papovich04}, reflecting
the difficulty of the `gastrophysics' that must be 
tackled to properly reproduce the full richness of 
the observational datasets.   For this reason, 
we focus in this section on qualitative
features of the models, rather than a detailed comparison 
of IR luminosity functions, or the distribution of
stellar masses and morphologies as a function of IR luminosity.

\citet{somerville01} present a `collisional starburst' model, 
in which major (with mass ratios $<$4:1) and minor (with mass
ratios between 4:1 and 10:1) mergers trigger an episode of rapid
star formation, with a star formation efficiency that increases
towards lower mass ratio (i.e., is higher for major mergers).  
At $z \sim 0$, \citet{somerville01} predict that 35\% of star formation 
is triggered by minor mergers, and 2.5\% of star formation is 
triggered by major mergers.  At $z \sim 0.7$, 50\% of star formation is 
triggered by minor mergers and 3.5\% by major mergers.  Thus, 
they predicted that the steep drop in the cosmic SFR since $z \sim 1$
is not caused by changes in the major merger rate; however, 
in their model a significant part of the drop reflects a 
declining rate of minor mergers.  There are a number of important
uncertainties: a lowered star formation efficiency in minor interactions
would shift more star formation into major mergers because more
gas would be available to form stars in major mergers, tidal 
interactions (e.g., flybys) are not included in the model, 
untriggered star formation in a galaxy with triggered star formation 
is not included in the census of triggered star formation, 
and our observational census of `clearly interacting' galaxies
may well include a significant fraction of minor mergers
and tidal interactions.

Nonetheless, it is encouraging that the observations and 
at least this model are reasonably consistent in the sense
that major galaxy mergers do not drive the declining 
cosmic SFR from $z \sim 1$ to the present day.  
At present, models cannot clearly predict which processes
dominate: while minor mergers and tidal interactions
may dominate \citep{somerville01}, it is also possible
that a dwindling gas supply could dominate \citep{cole00,somerville01}, 
depending on the efficiency of triggered star formation in the real universe.

\subsection{Improvements}

There are a number of areas that need to be improved
to more fully understand these results, and 
to illuminate the physics driving the cosmic SFH.  
We will mention
a few specific examples here. A more
complete, ground-based and IRAC near-IR-selected
photometric and/or spectroscopic
sample covering comparable or larger areas will be required to 
push our understanding past $z \sim 1$, into the regime
where we observe rapid cosmic star formation.  A wide range of 
morphological disturbance diagnostics must be developed and 
robustly tested \cite[see, e.g.,][for some promising 
examples of this kind of technique]{conselice03,lotz04}.
Total $L_{\rm IR}$ and SFR calibrations must be 
tested thoroughly using IR spectral data coupled with 
70{\micron} and 160{\micron}
imaging data where available.  This must be accompanied by a fuller analysis
of local, well-studied control samples to explore the fraction 
of $L_{\rm IR}$ coming from dust heated by optical light from 
old stellar populations, and possible metallicity dependences in 
IR SEDs.  The IR properties of high redshift AGN must be better
constrained, including constraints on the fraction of light 
emitted by AGN-heated dust compared with dust heated by associated
star formation in the AGN host.  Yet, the two main conclusions of this
paper --- the rapid demise of intermediate- and high-mass 
star-forming galaxies between 
$z \sim 0.7$ and the present-day, and the limited importance
of major mergers in driving the declining cosmic SFR --- are
likely to hold even after these studies are completed.

\section{Conclusions} \label{sec:conc}

In this paper, we have explored the 24{\micron}-derived
IR properties of an optically-selected sample of 1436 galaxies
with $0.65 \le z < 0.75$ in the CDFS.  Optical IDs, photometric
redshifts and morphologies were obtained from the COMBO-17 and GEMS 
surveys.  From examination of local LIRGs and ULIRGs, and exploration 
of COMBO-17's photometric redshift accuracy as a function of IR luminosity, 
we argue that the optical selection should not 
introduce any special bias against intensely star-forming
obscured galaxies.  Under the assumption that the mid-IR properties
of $z \sim 0.7$ galaxies are spanned by calibrating samples
in the local universe, we have derived total IR luminosities for
the 442 galaxies detected at better than $5 \sigma$ at 24{\micron}.
We used these data, in conjunction with the optical data, to 
estimate SFRs and stellar masses.  

We have found the following: 
\begin{itemize} 
\item Galaxies with X-ray-luminous 
AGN contribute $\la 15$\% towards the integrated $z \sim 0.7$
IR luminosity density.
\item The IR-to-UV ratio is a strong, increasing,
function of total SFR; this relation shows a $\ga 0.5$ dex scatter.  
There is no evidence for evolution 
of this relationship over the last 7 Gyr.  
\item Morphologically-normal galaxies show a strong relationship
between their optical and IR properties.  Optically 
blue early-type and spiral galaxies are IR-bright.  Optically
red early-type galaxies and a significant fraction of red spiral galaxies are
not detected at 24{\micron}.  Most galaxies on the
red sequence are genuinely old, red and dead.
\item Dust strongly shapes the optical properties of 
morphologically-peculiar galaxies.  The majority of obviously
interacting galaxies are detected at the star formation-sensitive
24{\micron} band, despite showing the full range of optical 
colors from very blue through to very red.
\item Clearly interacting galaxies with morphologies suggestive
of major galaxy mergers contribute at most 30\% 
of the integrated IR luminosity
density at $z \sim 0.7$.  Bearing in mind that there is a factor of three
reduction in cosmic IR luminosity density between $z \sim 0.7$
and the present day, this suggests that a declining major
merger rate can only form a small part of this drop in cosmic SFR.
\item Morphologically-normal galaxies (spirals, ellipticals 
and Magellanic irregulars) form more than 70\% of 
the integrated IR luminosity density in the $z \sim 0.7$ slice.  This implies
that physical processes which do not strongly affect galaxy morphology, 
such as gas consumption and weak tidal interactions with small satellite 
galaxies, are likely responsible for the drastic decline in cosmic SFR since 
$z \sim 0.7$.
\item At $z \sim 0.7$, nearly 40\% of intermediate- and high-mass galaxies are 
actively starbursting (i.e., they have $M_* \ge 2 \times 10^{10} M_{\sun}$ and 
$b \ge 1$).  In contrast, only 1\% of $z \sim 0$ 
galaxies with $M_* \ge 2 \times 10^{10} M_{\sun}$ are 
undergoing such active star formation.  The declining
SFR since $z \sim 0.7$ is driven by drastic evolution 
in the SFRs of intermediate- and high-mass 
galaxies, rather than starbursting dwarf galaxies.  
\end{itemize}

\acknowledgements

\vspace{0.8cm}

We wish to thank the referee, Jarle Brinchmann, for 
a constructive and thoughtful report which helped to improve
the paper.
This work is based (in part) on observations made 
with the Spitzer Observatory, which is operated by the Jet Propulsion
Laboratory, California Institute of Technology, under 
NASA Contract 107. Support for this work was 
provided by NASA through contract number 960785 issued by JPL/Caltech.
Support for the GEMS project was provided by NASA through grant number GO-9500
from the Space Telescope Science Institute, which is operated by the
Association of Universities for Research in Astronomy, Inc.\ for NASA, under
contract NAS5-26555.
E.\ F.\ B.\ was supported by the European
Community's Human Potential Program under contract
HPRN-CT-2002-00316, SISCO.
C.\ W.\ was supported by a PPARC Advanced Fellowship.
D.\ H.\ M.\ acknowledges support from NASA under LTSA Grant NAG5-13102 
issued through the Office of Space Science.
This publication made use of NASA's Astrophysics Data System 
Bibliographic Services.  This research has made use 
of the NASA/IPAC Extragalactic Database (NED) which is operated 
by the Jet Propulsion Laboratory, California Institute of Technology, 
under contract with the National Aeronautics and Space Administration.
This publication makes use of data products from the
{\it Two Micron All Sky Survey}, which is a joint project of the
University of Massachusetts and the Infrared Processing and
Analysis Center/California Institute of Technology, funded by
NASA and the National Science Foundation.

\appendix
\section{A. Are most starbursts in the COMBO-17 sample?}
\label{app:c17comp}

A possible source of bias could be 
that the most heavily-obscured $z \sim 0.7$ star-forming galaxies may be 
too optically faint to be reliably picked up and/or correctly 
classified by COMBO-17.  
Indeed, there are many MIPS sources without photometric 
redshifts ($m_R \ga 24$) or even detections ($m_R \ga 25.5$) from 
COMBO-17.  Without a comprehensive multi-wavelength photometric
redshift analysis we cannot definitively address this issue.
Yet, we explore this issue briefly here to assess
the potential scale of the biases, choosing to focus on 
two key issues: optical detectability and redshift accuracy.  

\begin{figure}[tb]
\begin{center}
\epsfxsize=9cm
\epsfbox{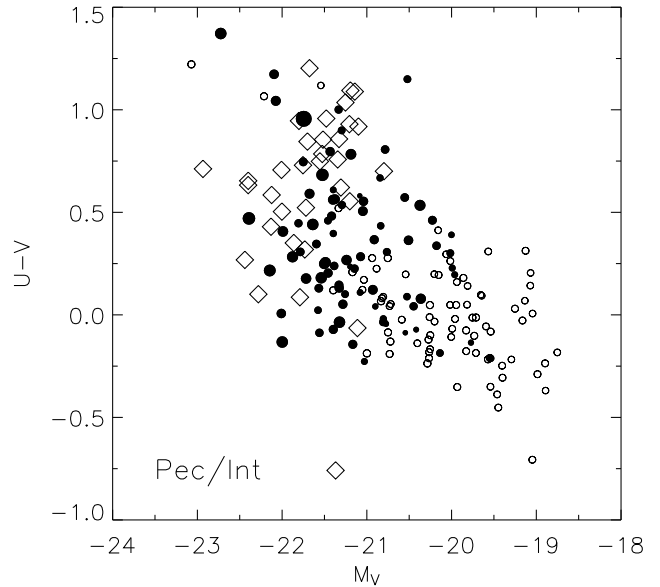}
\caption{\label{fig:ulirg}  
Rest-frame $U-V$ colors as a function of absolute magnitude 
in $V$-band for a sample of ULIRGs and LIRGs in 
the local universe (open diamonds).  Also plotted are
visually-classified peculiar and interacting galaxies
from the $z \sim 0.7$ slice, reproduced from Fig.\ \protect\ref{fig:cmr}.
 }
\end{center}
\end{figure}

One key point is that of optical brightness.  
In terms of affecting the total IR luminosity density or skewing 
the fractions of IR luminosity from each morphological type, the most
IR-luminous galaxies are the most relevant; i.e., luminous ($10^{12} L_{\sun} > L > 10^{11} L_{\sun}$; LIRGs) and ultra-luminous 
($> 10^{12} L_{\sun}$; ULIRGs) infrared galaxies.  
Thus, it is important to check 
if local ULIRGs and LIRGs would be visible if placed at $z \sim 0.7$.
We compare the optical properties of a sample
of local universe ULIRGs and LIRGs from \citet{surace00} and \citet{arribas04}
with the sample of Pec/Int galaxies from the $z \sim 0.7$ slice
(Fig.\ \ref{fig:ulirg}).   
Photometry for these galaxies was presented in $B$ and $I$ bands;
we corrected these for galactic foreground extinction, and 
transformed them to $U$ and $V$ absolute magnitudes using
the PEGASE stellar population model (if instead one assumed
a very young underlying stellar population, and postulated
that redder colors were due to dust alone, the transformation 
from $B-I$ to $U-V$ would remain unchanged to $\la 0.1$ mag, owing 
to the strength of the age/metallicity/dust degeneracy in the optical 
wavelength region).  In spite of their high
obscurations, local ULIRGs and LIRGs are reasonably bright in the optical,
and in common with $z \sim 0.7$ Pec/Int galaxies, show a wide 
range in galaxy colors (primarily because of the dust obscuration). 
Highly-obscured IR-luminous 
galaxies are {\it not} faint in the optical; they are
rather fainter than they would be in the absence of dust, but they 
are still reasonably bright, and are certainly bright enough to be 
easily picked up at $z \sim 0.7$ to an $m_R$-band limit of 
24\footnote{ULIRGs and LIRGs at $z \ga 1$ will 
have $m_R \ga 24$; therefore,
optically-selected redshift surveys 
to this limit will be unable to construct 
unbiased samples of IR-luminous galaxies.}.

\begin{figure}[tb]
\begin{center}
\epsfxsize=9cm
\epsfbox{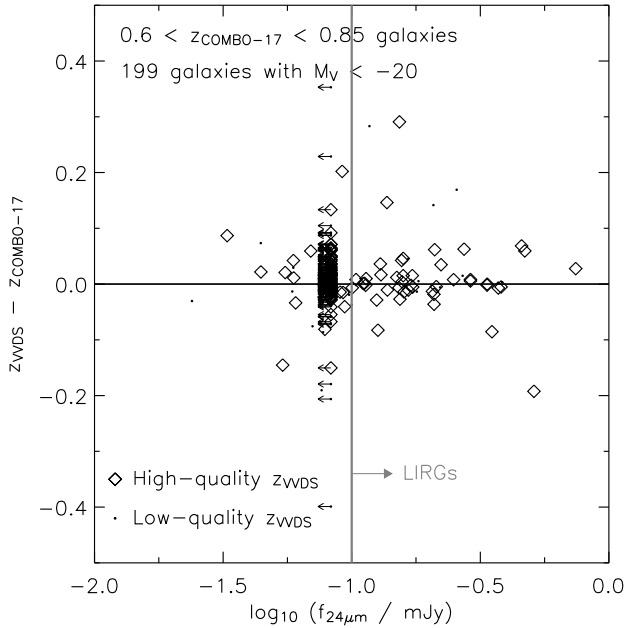}
\caption{\label{fig:ztest}  
A comparison of VVDS spectroscopic and COMBO-17 photometric redshifts
as a function of 24{\micron} flux.  Diamonds denote high-quality
spectroscopic redshifts and points denote those with more
uncertain line identifications.  199 galaxies with $M_V < -20$ and 
$0.6 < z < 0.85$ are shown.  The grey vertical line approximately shows the 
canonical $10^{11} L_{\sun}$ limit for LIRGs, showing a considerable LIRG
population in this field and redshift range.  Arrows denote 24{\micron}
upper limits.
 }
\end{center}
\end{figure}

Yet, it is not clear that these dusty galaxies will be well-classified 
by COMBO-17's photometric redshift classifier.  COMBO-17 adopts
a 2-dimensional stellar population and SMC extinction curve
dust reddening grid for classification, allowing it in principle
to classify galaxies with a wide diversity of stellar population
properties and dust contents.  In Fig.\ \ref{fig:ztest}
we test the performance of COMBO-17's classifier directly
using redshifts from the VIMOS VLT Deep Survey \citep[VVDS;][]{lefevre04}
as a function of the 24{\micron} flux.  To ensure a large sample,
we expand the redshift range to $0.6 < z_{\rm phot} < 0.85$,
and we require that $M_V < -20$, as this range contains the vast
majority of all MIPS detections.  There are 
a total of 199 galaxies with VVDS redshifts with estimated
photometric redshifts and rest-frame $V$-band magnitudes 
in this range.  Diamonds show galaxies
with well-measured spectroscopic redshifts (quality flags 3/4/23/24),
whereas lower-quality spectroscopic redshifts, typically with less certain
line identifications, are shown as 
points (quality flags 1/2/9/21/22).  It is clear that most  
galaxies have well-constrained photometric redshifts; the robust
sigma of the distribution is $\sigma_{\rm robust} (\Delta z) = 0.04$, 
independent of IR luminosity.  Furthermore, 
the fraction of LIRGs with $\Delta z > 0.2$ is 5/57, or 9\%, whereas
the fraction of galaxies with $L < 10^{11} L_{\sun}$ with 
$\Delta z > 0.2$ is 7/142, or 5\%\footnote{The worst outliers
have lower quality VVDS spectroscopic redshifts; Wolf (priv. comm.) 
finds in a more comprehensive comparison that the VVDS spectroscopic
redshifts with low confidence 
quality flags 1/2/9/21/22 have a $\sim 25$\% error
rate, and that COMBO-17 has an overall error rate of $\sim 5$\% or less.}.  
The RMS and outlier fraction do not strongly depend on IR luminosity.

Thus, bearing in mind that $z \sim 0.7$ IR-luminous galaxies 
are reasonably bright in the optical, and that COMBO-17's classifier
appears to work equally well for IR-bright and IR-faint galaxies, we
tentatively conclude that the philosophy used in this paper --- exploring
the IR properties of an optically-selected sample in a thin redshift
slice --- should not lead to any serious biases against IR-luminous, 
heavily obscured galaxies.

\vspace{1.5cm}

\section{B. A comparison of GOODS and GEMS morphologies of IR-bright galaxies}
\label{app:goods}

An unavoidable source of error comes from classification 
uncertainties.  With the classifications based on the 
1-orbit rest-frame $V$-band data from GEMS, $\ga 70$\% of 
the SFR is in undisturbed galaxies.  Yet, when one probes
to fainter surface brightness levels, one will be able to 
better recognize tidal tails, leading to an increased
fraction of interacting galaxies.
To estimate the importance of this source of uncertainty, 
we classified a subset of 290 $0.65 \le z < 0.75$ galaxies
with deeper F850LP data from GOODS 
\citep[][the GOODS southern dataset forms the central $\sim 1/4$ of the GEMS 
area, giving F850LP data with 5$\times$ longer exposure time]{giavalisco04}. 
Overall, around twice as many galaxies
appeared to be clearly interacting
in the deeper dataset; the detailed increase
in fraction varied from classifier to classifier.
Almost all of the newly-recognized interactions were
relatively faint --- GEMS-depth data are insufficient to 
unambiguously discern the faint tidal features that 
are visible in the deeper GOODS data.  This issue is 
discussed at length in \citet{wolfuv}.

Owing to small number statistics, it is impractical to 
simply analyze the properties of the GOODS sample alone.
We therefore derive a correction to the IR luminosity
function that is tuned to reproduce the systematic difference
between the GOODS and GEMS classifications for the 290-galaxy
subsample covered by GOODS.  For each galaxy, the nearest
25 galaxies in terms of IR luminosity are selected.  Then 
relative split of spiral : E/S0 : Irr : Pec/Int 
for these 25 galaxies using the GEMS classifications is derived, and 
is compared with the type split derived using 
the GOODS classifications.  This process is repeated for all 
galaxies, eventually building up the split of 
galaxy types as a function of IR luminosity in 
GEMS and GOODS.  The difference between
these type splits, as a function of IR luminosity, 
is used to correct the GEMS type-dependent IR 
luminosity functions in Fig.\ \ref{fig:lf} to 
roughly compensate for the shallower depth of GEMS
compared to GOODS.  
The main differences between GOODS and GEMS is that GOODS-depth
data shows a statistically-significant increased fraction of 
clear interactions at $\log_{10} L_{\rm IR}/L_{\sun} \sim 11$, 
while the fraction of irregular and spiral galaxies at these luminosities
is slightly reduced to compensate.  

This issue will be discussed further by Papovich et al.\ (in preparation).

\end{document}